\documentclass[conference]{IEEEtran}
\IEEEoverridecommandlockouts
\usepackage{cite}
\usepackage{amsmath,amssymb,amsfonts}
\usepackage{algorithmic}

\usepackage{graphicx}
\usepackage{textcomp}
\usepackage{xcolor}
\def\BibTeX{{\rm B\kern-.05em{\sc i\kern-.025em b}\kern-.08em
    T\kern-.1667em\lower.7ex\hbox{E}\kern-.125emX}}
\usepackage{graphicx}
\usepackage{subcaption}
\graphicspath{ {images/} }
\usepackage[utf8]{inputenc}
\usepackage{fancyhdr}
\usepackage{dirtytalk}
\usepackage{tabularx}
\PassOptionsToPackage{hyphens}{url}\usepackage{hyperref}
\usepackage[ruled,lined,linesnumbered]{algorithm2e}

\usepackage{enumitem,kantlipsum}

\usepackage{graphicx}
\graphicspath{ {images/} }
\usepackage[utf8]{inputenc}
\usepackage{fancyhdr}
\usepackage{dirtytalk}
\usepackage{tikz}
\usetikzlibrary{shapes.geometric, arrows}
\usepackage{pgfplots}
\pgfplotsset{width=8cm,compat=1.16}
\usepgfplotslibrary{external}
\usepackage{mathtools}
\usepackage{multirow}
\usepackage{booktabs}
\usepackage{epstopdf}
\usepackage{setspace} 

\tikzstyle{startstop} = [rectangle, rounded corners, minimum width=2cm, minimum height=0.5cm,text centered, draw=black, fill=red!30]
\tikzstyle{process} = [rectangle, minimum width=2cm, minimum height=0.5cm, text centered, draw=black, fill=orange!30, align=left]
\tikzstyle{decision} = [diamond, minimum width=1.0cm, minimum height=0.4cm, text centered, draw=black, fill=green!30]
\tikzstyle{arrow} = [thick,->,>=stealth]

\newcolumntype{L}{>{$}l<{$}}

\usepackage[labelformat=simple]{subcaption}

\begin{document}


\title{Contention-Aware Microservice Deployment in Collaborative Mobile Edge Networks\thanks{This work is supported by the National Science Foundation of China under Grant 62271062.}}

\author{\IEEEauthorblockN{Xinlei Ge, Yang Li, Xing Zhang\IEEEauthorrefmark{1}, Yukun Sun, Yunji Zhao}

\IEEEauthorblockA{ Wireless Signal Processing and Network Laboratory \\
Beijing University of Posts and Telecommunications, Beijing 100876, China\\
\IEEEauthorrefmark{1} E-mail: zhangx@ieee.org}
}

\maketitle
\begin{abstract}
As an emerging computing paradigm, mobile edge computing (MEC) provides processing capabilities at the network edge, aiming to reduce latency and improve user experience. Meanwhile, the advancement of containerization technology facilitates the deployment of microservice-based applications via edge node collaboration, ensuring highly efficient service delivery. However, existing research overlooks the resource contention among microservices in MEC. This neglect potentially results in inadequate resources for microservices constituting latency-sensitive applications, leading to increased response time and ultimately compromising quality of service (QoS). To solve this problem, we propose the Contention-Aware Multi-Application Microservice Deployment (CAMD) algorithm for collaborative MEC, balancing rapid response for applications with low-latency requirements and overall processing efficiency. The CAMD algorithm decomposes the overall deployment problem into manageable sub-problems, each focusing on a single microservice, then employs a heuristic approach to optimize these sub-problems, and ultimately arrives at an optimized deployment scheme through an iterative process. Finally, the superiority of the proposed algorithm is evidenced through intensive experiments and comparison with baseline algorithms.
\end{abstract}


\section{Introduction}
With the widespread adoption of IoT devices and the growth of mobile traffic, the demand for low-latency and high-performance computing has surged. To mitigate the delay issues in long-distance data transmission in traditional cloud computing, mobile edge computing (MEC) servers with computational and storage capabilities are positioned in proximity to users\cite{yukun2024computing}. Microservice architecture is characterized by decomposing large applications into loosely coupled and function-specific microservices; this approach has become predominant in modern software design due to its ability to enable rapid delivery and scale\cite{jamshidi2018microservices}. Advances in containerization technology resolve the integration issues between microservices and operating systems\cite{alqaisi2023containerized}, facilitating their deployment on edge servers that can communicate with each other, which in turn enables collaborative processing of application requests. However, in contrast to cloud computing, MEC faces resource constraints\cite{li2024priority}, causing resource competition among applications with diverse latency requirements, which is intrinsically the contention among microservices. Inadequately designed deployment schemes lead to suboptimal resource allocation, resulting in delayed response time to urgent tasks. In addition, the complex dependencies between microservices increase the challenges of deployment scheme design.

In recent years, there is growing interest in studying microeservice deployment in MEC. In some studies considering dependencies, each microservice is restricted to running on a single service node. For example, the authors of \cite{qi2023edge} proposed a distributed deep reinforcement learning algorithm to reduce response latency and balance resource utilization. \cite{zhou2023dependency} introduced a coalition game-based algorithm to devise the microservice deployment strategy while accounting for dependencies. In \cite{li2023topology}, the authors proposed a topology-aware framework that utilizes a graph mapping algorithm to reduce network overhead and improve resource utilization.

Additionally, some literature focuses on microservice deployment within a single application. In \cite{deng2020optimal}, the authors employed a branch-and-bound-based approach to address the microservice deployment challenge, aiming to meet QoS requirements while minimizing deployment cost. Some studies explore deployment for multiple applications; however, they fail to sufficiently investigate the resource competition among microservices from applications with different latency requirements during deployment. For example, the authors of \cite{li2022application} studied a microservice deployment method based on the genetic algorithm to achieve low latency and load balancing, but ignored the resource competition problem.

However, constraining the deployment of each microservice to a single node significantly impairs load balancing and latency optimization. Moreover, user requests are usually not limited to one type of application, and resource contention issues cannot be ignored. Therefore, we study the problem of deploying microservices of multiple applications with different latency requirements in collaborative MEC, thoroughly considering resource contention in the system. To obtain the deployment scheme, we propose the Contention-Aware Multi-Application Microservice Deployment (CAMD) algorithm. The main contributions are summarized as follows:
\begin{itemize}
	\item We formulate the microservice deployment problem which aims at minimizing response time while accounting for resource competition among microservices. This optimization problem falls into the category of NP-hard problems.
	
	\item To solve this problem, we design the CAMD algorithm based on the idea of the Block Coordinate Descent (BCD) algorithm. The overall deployment scheme is decomposed into individual sub-schemes for each microservice, and then a heuristic algorithm is applied to optimize the sub-schemes.
	
	\item Sufficient experiments are conducted to verify the effect of the proposed algorithm using both Python and a Kubernetes-based testbed. Experimental results show that CAMD outperforms baseline algorithms in reducing response time.
\end{itemize}
\section{System Model}
\subsection{Servers and Network}
\begin{figure}[ht!]
    \centering
    \includegraphics[width=0.4\textwidth]{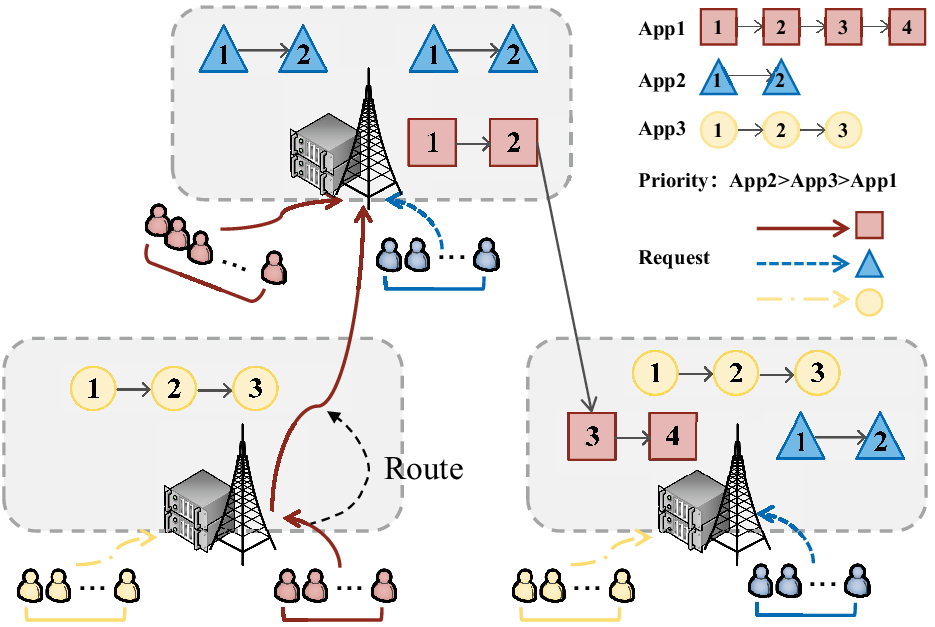}
    \caption{System model.}
    \label{fig:conf_system_model}
\end{figure}

Fig.~\ref{fig:conf_system_model} illustrates our MEC system, which consists of several end users and a set of edge servers, each co-located with its corresponding base station. Requests from users are first sent to the nearest base station and immediately forwarded to the co-located edge server for processing. The set of servers is symbolized as $\mathcal{S}=\{s_1,...,s_i,s_j,...,s_{|\mathcal{S}|}\}$. $s_i$ and $s_j$ can transfer data between each other, with the bandwidth denoted as $b_{s_i,s_j}$, where $b_{s_i,s_j}$ belongs to the bandwidth set $\mathcal{B}$. Due to differences in hardware specifications and storage capacity, different edge servers exhibit computational heterogeneity. The CPU and Memory capacities of different servers can vary, denoted as $CPU_{s_i}$ and $MEM_{s_i}$. 
\vspace{-10pt} 
\subsection{Application Model}
The MEC system is designed to support $|\mathcal{A}|$ different types of applications. $\mathcal{A}=\{a_1,...,a_k,...,a_{|\mathcal{A}|}\}$ represents the set of applications. Application priorities correlate with latency demands, where higher latency sensitivity yields higher priority. Each application $a_k$ is assigned a relative priority $\gamma_{a_k}$, and these priorities are normalized so that they sum to unity across all applications, i.e., $\sum_{k=1}^{|\mathcal{A}|} \gamma_{a_k} = 1$. $|a_k|$ represents the number of microservices that comprises $a_k$. An application $a_k$, consisting of dependent microservices, can be represented as a Directed Acyclic Graph (DAG) $G_{a_k} = <V_{a_k}, E_{a_k}>$, where the vertex set $V_{a_k}=\{ms_1^{a_k},...,ms_v^{a_k},...,ms_{|a_k|}^{a_k}\}$ corresponds to the microservices and the edge set $E_{a_k}=\{e_{1,2}^{a_k},e_{2,3}^{a_k},...,e_{|a_k|-1,|a_k|}^{a_k}\}$ captures the dependency relationships among these microservices. If $e_{v,v+1}^{a_k} = 1$, it signifies that $ms_{v+1}^{a_k}$ is executed subsequent to $ms_{v}^{a_k}$, relying on its computational output, with the data transfer size between them denoted as $w_{v,v+1}^{a_k}$. 

Microservices run as separate instances with specific CPU and Memory resource requirements to ensure optimal performance and efficiency. For each instance of $ms_v^{a_k}$, $cpu_v^{a_k}$ and $mem_v^{a_k}$ represent the required CPU and Memory resources respectively. $c_{ms_v^{a_k}}$ is the CPU cycles required by $ms_v^{a_k}$ to process a single incoming request. The processing rate of a microservice instance is defined as $\frac{cpu_v^{a_k}}{c_{ms_v^{a_k}}}$, denoted as $o_v^{a_k}$. Like several other studies including \cite{wang2022online}, \cite{chen2023msm} and \cite{li2023joint}, we also divide time into discrete slots. In time slot $t$, the total number of requests for $a_k$ in the system is denoted as $R^{a_k}(t)$. 
\subsection{Processing Path}
User requests for $a_k$ are sent to the nearest base station, whereupon the co-located server receives these requests. Subsequently, requests are processed sequentially by microservice instances of $a_k$, which are potentially distributed across different servers. The processing path is defined as the sequence of servers traversed by requests from reception to completion of processing. $p(ms_v^{a_k}) = s_i$ denotes that requests for $ms_v^{a_k}$ are handled by instances on $s_i$, where $1 \leq v \leq |a_k|$. A processing path of $a_k$ is modeled as $\mathcal{P}_m^{a_k}=\{p(ms_0^{a_k})$$,p(ms_1^{a_k})$$,p(ms_2^{a_k})$$,...,p(ms_{|a_k|}^{a_k})\}$, with $p(ms_0^{a_k})$ representing the server receiving the requests. $P_{\mathcal{P}_m^{a_k}}(t)$ denotes the probability of the processing path being $\mathcal{P}_m^{a_k}$ in time slot $t$. The expression for the response latency of requests for $a_k$ is as follows:
{\small \begin{equation}
	T_{a_k}(t) = \sum \limits _{p(ms_0^{a_k})=s_1} ^{s_{|\mathcal{S}|}} \sum \limits _{p(ms_1^{a_k})=s_1} ^{s_{|\mathcal{S}|}}... \sum \limits _{p(ms_{|a_k|}^{a_k})=s_1} ^{s_{|\mathcal{S}|}} P_{\mathcal{P}_m^{a_k}}(t)T_{\mathcal{P}_m^{a_k}}(t)
\end{equation}}
where {$\!T_{\mathcal{P}_m^{a_k}}$$(t)$} is the response latency of $\mathcal{P}_m^{a_k}$. $P_{\mathcal{P}_m^{a_k}}$$(t)$ is expressed as 
{\small\begin{equation}
	P_{\mathcal{P}_m^{a_k}}(t)=P_{p(ms_0^{a_k})}(t)\prod_{v=1}^{|a_k|}P_{ms_v^{a_k},p(ms_v^{a_k})}(t)
\end{equation}}
where $P_{p(ms_0^{a_k})}(t)$ represents the probability that the requests are first sent to the server $p(ms_0^{a_k})$, and $P_{ms_v^{a_k},p(ms_v^{a_k})}(t)$ denotes the probability that requests for $ms_v^{a_k}$ are processed on the server $p(ms_v^{a_k})$. $P_{p(ms_0^{a_k})}(t)$ is determined by the distribution of user requests. The request number of $a_k$ received on $s_i$ is denoted as $r_{ p(ms_0^{a_k})= s_i}^{a_k}(t)$, with $r_{s_i}^{a_k}$ being an element of the request set $\mathcal{R}$.  $P_{p(ms_0^{a_k})}(t)$ is calculated as: 
{\small\begin{equation}
	P_{p(ms_0^{a_k})=s_i}(t)=\frac {r_{ p(ms_0^{a_k})= s_i}^{a_k}(t)} {R^{a_k}(t)}
\end{equation}}
{\small\begin{equation}
	R^{a_k}(t) = \sum \limits _{i=1} ^{|\mathcal{S}|} r_{p(ms_0^{a_k})=s_i}^{a_k}(t)
\end{equation}}

Each microservice usually deploys multiple instances distributed across different servers. The selection of a specific server to handle the incoming requests depends on the routing policy. There are various rooting policies, such as round-robin policy, weighted routing policy and random policy. This paper adopts the round-robin policy, which is a commonly used strategy that provides equal opportunities for each instance in terms of receiving requests. Under this scheduling strategy, the probability of a server processing requests for $ms_v^{a_k}$ is proportional to the instance number of $ms_v^{a_k}$ deployed on the server.
In time slot $t$, $P_{ms_v^{a_k},p(ms_v^{a_k})}(t)$ is formulated as
{\small\begin{equation}
	P_{ms_v^{a_k},p(ms_v^{a_k})=s_i}(t)=\frac {n_{ms_v^{a_k},p(ms_v^{a_k})=s_i}(t)}{N_{ms_v^{a_k}}(t)}
\end{equation}
\begin{equation}
	N_{ms_v^{a_k}}(t)=\sum \limits _{i=1}^{|\mathcal{S}|} n_{ms_v^{a_k},p(ms_v^{a_k})=s_i}(t)
\end{equation}}
where the total number of instances of $ms_v^{a_k}$ is $N_{ms_v^{a_k}}(t)$, with $n_{ms_v^{a_k},s_i}(t)$ denoting the instance number of $ms_v^{a_k}$ deployed on $s_i$.
\vspace{-0.268cm}
\subsection{Request Response Time}
In this paper, the transmission delay of user requests to the base station is not affected by the deployment of microservices, thus it is ignored. We also disregard the results transmission time to end users, as the small data volume renders this duration negligible. The formula for response latency is given by (7).
\begin{equation}
	T_{\mathcal{P}_m^{a_k}}(t)=T_{\mathcal{P}_m^{a_k}}^{tran}(t)+T_{\mathcal{P}_m^{a_k}}^{com}(t)
\end{equation}where $T_{\mathcal{P}_m^{a_k}}^{tran}(t)$ and $T_{\mathcal{P}_m^{a_k}}^{com}(t)$ represent data transmission latency between microservices and computing delay, respectively. $T_{\mathcal{P}_m^{a_k}}^{tran}(t)$ is taken into consideration when the server initially receiving the requests differs from the server executing the first microservice, or when the actual processing instances of adjacent microservices are located on different servers. $T_{\mathcal{P}_m^{a_k}}^{tran}(t)$ is expressed as follows: 
{\small\begin{equation}
\begin{split}
	T_{\mathcal{P}_m^{a_k}}^{tran}(t)=
\begin{cases} 
	0, & p(ms_0^{a_k})= p(ms_1^{a_k}) \\
	\frac{w_{req}^{a_k}}{b_{p(ms_0^{a_k}),p(ms_1^{a_k})}}, & p(ms_0^{a_k})\neq p(ms_1^{a_k})
\end{cases} 
\\ + \sum \limits _{v=1}^{|a_k|-1}
\begin{cases} 
	0, & p(ms_v^{a_k})= p(ms_{v+1}^{a_k}) \\
	\frac{w_{ms_{v,v+1}^{a_k}}}{b_{p(ms_v^{a_k}),p(ms_{v+1}^{a_k})}}, & p(ms_v^{a_k})\neq p(ms_{v+1}^{a_k}) 
\end{cases} 
\end{split}
\end{equation}}where $w_{req}^{a_k}$ is the size of the request data for $a_k$.

$T_{\mathcal{P}_m^{a_k}}^{com}(t)$ can be defined as the accumulated sum of computing time $t_{ms_v^{a_k}}$ of all microservices of $a_k$. $T_{\mathcal{P}_m^{a_k}}^{com}(t)$ can be computed as follows:
\begin{equation}
	T_{\mathcal{P}_m^{a_k}}^{com}(t)=\sum \limits _{v=1}^{|a_k|}t_{ms_v^{a_k}}=\sum \limits _{v=1}^{|a_k|} \frac{R^{a_k}(t)}{N_{ms_v^{a_k}}(t)} \times \frac{1}{o_v^{a_k}}
\end{equation}
\subsection{Deployment Scheme}
To maintain system stability, the deployment scheme is updated at a limited frequency of once per time slot. In the remainder of this paper, we focus on the deployment scheme in one time slot and the time parameter $t$ will be omitted in our notation. 
The deployment scheme is defined as $\mathcal{N}$$ =$$ \{$$sc_{ms_v^{a_k}}$$|$$a_k$$ \in$$ \mathcal{A}$$, $$ms_v^{a_k} $$\in$$ V_{a_k}$$\}$, with $sc_{ms_v^{a_k}}$$=$$\{$$n_{ms_v^{a_k},s_1},$$n_{ms_v^{a_k},s_2}$,$\dots$,$n_{ms_v^{a_k},s_{|\mathcal{S}|}}$$\}$.
\subsection{Problem Formulation}
In this paper, our optimization objective is to minimize the request response time, which is defined as the weighted sum of response time for all application requests, where the weight for each application is determined by its priority. This formulation ensures that reduction in high-priority application response time contributes more to the optimization goal, promoting an optimal balance between responsiveness and resource allocation. The optimization problem is expressed as follows:
{\small\begin{equation}
\begin{split}
	\mathcal{P}_1:& min \sum \limits _{k=1}^{|\mathcal{A}|} \gamma_{a_k}T_{a_k}
	\\&(a) \sum \limits _{k=1}^{|\mathcal{A}|} \sum \limits _{v=1}^{|a_k|} n_{ms_v^{a_k},s_i} \times cpu_v^{a_k} \leq CPU_{s_i}, \forall s_i \in \mathcal{S} \\
	&(b)\sum \limits _{k=1}^{|\mathcal{A}|} \sum \limits _{v=1}^{|a_k|} n_{ms_v^{a_k},s_i} \times mem_v^{a_k} \leq MEM_{s_i}, \forall s_i \in \mathcal{S} \\
	&(c)\sum \limits _{i=1}^{|\mathcal{S}|}n_{ms_v^{a_k},s_i} \geq 1
\end{split}
\end{equation}}

In the above formulation, (a) and (b) represent the CPU and Memory restrictions for the servers, respectively. (c) indicates that each microservice deploys at least one instance to ensure the normal operation of each application. The optimization problem is an NP-hard problem, which results in high computational complexity, making direct solutions infeasible. In order to effectively solve this problem, we propose the CAMD algorithm, which is described in detail in the next section.
\section{Microservice Deployment Algorithm}
To enable quick handling of time-critical applications without compromising system-wide efficiency, we propose the CAMD algorithm, a tri-phase approach that tackles $\mathcal{P}_1$. Firstly, the instance number of each microservice can be derived from application priorities and request volumes. In the microservice instances deployment phase, inspired by the BCD algorithm, we decompose the overall problem into multiple sub-problems, each focusing on a specific microservice deployment decision, with each sub-problem solved by the simulated annealing (SA) algorithm. Finally, the deployment scheme is adjusted through migration to ensure server resource limits are not exceeded. A detailed analysis of these three phases is presented in the following sections.
\vspace{-0.15cm}
\subsection{Instance Count Calculation}
\textbf{Step1: }In resource allocation, applications with high request volumes require more computing resources to avoid increased response time. Simultaneously, time-critical applications require guaranteed resource allocation to maintain their QoS across varying levels of request volumes. Considering these factors, resource allocation is determined by the priority and request volume of each application. Resource allocation is quantified by the number of chains per application, denoted as $Num_{a_k}$, which is proportional to the product of the priority and request count of the application.
{\begin{equation}
		\begin{split}
			Num_{a_1}:Num_{a_2}:...:Num_{a_{|\mathcal{A}|}} \\= R^{a_1} \times \gamma_{a_1}:R^{a_2} \times \gamma_{a_2}:...:R^{a_{|\mathcal{A}|}} \times \gamma_{a_{|\mathcal{A}|}} 
		\end{split}
\end{equation}}
\indent \textbf{Step2: }Within each application, the number of microservice instances should be inversely proportional to the processing rate of each microservice (12). Microservices with a relatively lower processing rate should be allocated more instances, as they often become system bottlenecks resulting in overall performance degradation. Deploying more instances of these microservices can effectively alleviate their processing pressure.
\begin{equation}
	\begin{split}
		N_{ms_1^{a_k}}:N_{ms_2^{a_k}}:...:N_{ms_{|a_k|}^{a_k}} \\= \frac {1}{o_1^{a_k}}:\frac {1}{o_2^{a_k}}:...:\frac {1}{o_{|a_k|}^{a_k}} 
	\end{split}
\end{equation}

\textbf{Step3: }Based on (11), (12) and the total available resources, for every application, the specific instance number of each microservice can be calculated by solving the following equations (13) and (14).
\begin{equation}
	\sum \limits _{k=1}^{|\mathcal{A}|} Num_{a_k}^{'} \sum \limits _{v=1}^{|a_k|} N_{ms_v^{a_k}}^{'}\times cpu_v^{a_k} = \sum \limits _{i=1}^{|\mathcal{S}|} CPU_{s_i}
\end{equation}
\begin{equation}
	\sum \limits _{k=1}^{|\mathcal{A}|} Num_{a_k}^{''} \sum \limits _{v=1}^{|a_k|} N_{ms_v^{a_k}}^{''}\times mem_v^{a_k} = \sum \limits _{i=1}^{|\mathcal{S}|} MEM_{s_i}
\end{equation}

After solving (13) and (14), we can get two sets of solutions $\mathcal{NUM}^{'}=\{N_{ms_v^{a_k}}^{'}|a_k \in \mathcal{A}, ms_v^{a_k} \in V_{a_k}\}$ and $\mathcal{NUM}^{''}=\{N_{ms_v^{a_k}}^{''}|a_k \in \mathcal{A}, ms_v^{a_k} \in V_{a_k}\}$. The smaller one is taken as the initial deployment number of microservice instances, denoted as $\mathcal{NUM}^{*}=\{N_{ms_v^{a_k}}^{*}|a_k \in \mathcal{A}, ms_v^{a_k} \in V_{a_k}\}$.

\textbf{Step4:}
The instance number of each microservice is determined in Step 3. These instances are randomly deployed across the available servers to generate the initial deployment scheme $\mathcal{N}^{*}$, $\mathcal{N}^{*}=\{sc_{ms_v^{a_k}}^{*}|a_k \in \mathcal{A}, ms_v^{a_k} \in V_{a_k}\}$, where $sc_{ms_v^{a_k}}^{*}=\{n_{ms_v^{a_k},s_1}^{*},n_{ms_v^{a_k},s_2}^{*},...,n_{ms_v^{a_k},s_{|\mathcal{S}|}}^{*}\}$ satisfies the condition that $n_{ms_v^{a_k},s_1}^{*}+n_{ms_v^{a_k},s_2}^{*}+...+n_{ms_v^{a_k},s_{|\mathcal{S}|}}^{*}=N_{ms_v^{a_k}}^{*}$.
\subsection{Instance Deployment}
In this phase, instances of each microservice are deployed to the servers according to the specified number $N_{ms_v^{a_k}}^{*}$, as outlined in Algorithm 1. The deployment algorithm framework follows the BCD algorithm idea, tackling multi-variable optimization by optimizing different subsets of variables in each iteration, while temporarily keeping the others fixed. The algorithm steps include:
$\langle1\rangle$ Block Selection: In BCD, the first step is to determine how to group the variables. In this paper, each iteration optimizes the deployment of a single microservice while keeping others fixed (line 4).
$\langle2\rangle$ Block Optimization: We optimize each block using the SA algorithm. To implement the SA algorithm, we set $\mathcal{N}^{*}$ as the initial solution and (10) as the objective function. In this paper, the neighborhood function is defined as a swap operation, i.e., selecting two microservice instances and exchanging their servers (lines 9-10). This process (lines 7-18) is repeated until the termination condition is met.
$\langle3\rangle$ Inter-Block Iteration: Upon completing the deployment optimization of a microservice, proceed to the next one and repeat the process. The order is based on the priority of the application, from high to low, with microservices of the same application optimized in the order of the DAG. 
$\langle4\rangle$ Global Iteration: Once all microservices are optimized in sequence (lines 4-19), the whole process can be repeated multiple times until the stopping condition is met - either reaching the maximum number of iterations (line 2) or detecting no further changes in the deployment scheme (line 21).
\begin{algorithm}[ht!]
\SetAlgoLined
\SetKwInOut{Input}{Input}
\SetKwInOut{Output}{Output}
\Input{$\text{Application set: }\mathcal{A}, \text{Server set: } \mathcal{S}, \text{Request set: } \mathcal{R},$\par $\text{Bandwidth set: } \mathcal{B} ,\text{Initial deployment scheme: }\mathcal{N}^{*}$}
\Output{$\text{Deployment scheme:}$  $\mathcal{N}'$}
 \SetKwBlock{Beginn}{beginn}{ende}
$\mathcal{SC}' \gets \mathcal{N}^*,\mathcal{SC}'' \gets \emptyset, iter \gets MAX$ \\
    \While{$iter > 0$}{
    	$\mathcal{SC}'' \gets copy(\mathcal{SC}')$\\
    	\For{$sc_{ms_{v}^{a_k}}^{*} \in \mathcal{SC}'$}{
    		 $State \gets \mathcal{SC}'\backslash sc_{ms_{v}^{a_k}}^{*},$ $T^{'} \gets T_{initial},$
    		  $curT \gets calT(\mathcal{SC}',\mathcal{A},\mathcal{S},\mathcal{R},\mathcal{B})$\\
    		
    		$calT$: Calculate the response time based on the formula $\sum \limits _{k=1}^{|\mathcal{A}|} \gamma_{a_k}T_{a_k}$\\
    		\While{$T^{'} > T_{min}$}{
    			\For{$i=1,\ldots,L$}{
    			$\text{Select } s_i$ and $s_j$ that deploy instances of $ms_v^{a_k}$ \\
    			$newState \gets sc_{ms_{v}^{a_k}}^{*}\{n_1,...,n_i \gets n_i + 1,$ $...,$ $ n_j$ $\gets$ $n_j - 1$,$...\} $ \\
    			$newT$ $\gets$ $calT$($State$$\cup$$\{newState\}$$,$$\mathcal{A},$$\mathcal{S},$$\mathcal{R},$$\mathcal{B}$)\\
    			$\Delta E \gets newT - curT$ \\
    			\If{$\Delta E < 0 || \exp{(-\Delta E / T^{'})} > random $}{$sc_{ms_{v}^{a_k}}^{*} $$\gets$$ newState,$ $curT \gets newT$}
    		}
    			$T^{'} \gets T^{'} * \alpha $
    		}
    	}
    	$iter \gets iter-1$ \\
    	\If{$\mathcal{SC}' == \mathcal{SC}''$}{
    		$break$
    	}
    }
$\mathcal{N}' \leftarrow \mathcal{SC}'$ \\
\textbf{return} $\mathcal{N}'$
 \caption{BCD-Based Instance Deployment Algorithm}
 \label{algo:feedback_filtering}
\end{algorithm}
\vspace{-0.5cm}
\subsection{Instance Migration}
In this phase, the infeasible solution $\mathcal{N}^{'}$ is converted to a feasible one $\mathcal{N}$. The deployment of each microservice instance is evaluated, and if an instance causes the server to exceed the resource limits, the system checks for a server with enough resources. If available, the instance is migrated; if not, the instance is removed.
\section{Performance Evaluation}
This section begins with an overview of the experimental setup, followed by a comprehensive presentation of the results.
\subsection{Experiment Configuration}
In this paper, the total CPU frequency and memory capacity of each server are distributed in [5-20] GHz and [80-640] GB, respectively. The bandwidth between servers is set to 1$\pm$0.2 Gbps. For each microservice, we set the required CPU frequency to [0.1-0.5] GHz, the required CPU cycles per request to [2.4-12] M cycles, along with the memory requirement of [0.5-4] GB. The amount of data transmitted between two microservices varies from 1 KB to 100 KB. 

The following algorithms serve as baselines for comparison with the proposed algorithm:

\begin{itemize}
	 \item \textbf{DeploySpread:} The DeploySpread\cite{he2022online} algorithm is based on the greedy algorithm. When deploying a new instance of $ms_v^{a_k}$, it selects the server with sufficient resources and minimal latency, considering the impact on related microservices. After the deployment, the instances of the predecessors and successors of $ms_v^{a_k}$ are redeployed.
	
	\item \textbf{MOCP-MFEA:} The deployment algorithm used in \cite{liu2021multi} is based on the multi-objective evolutionary algorithm. In this paper, we adjust the optimization objective to response latency.
	
	\item \textbf{Gurobi:} First, the instance number of $ms_v^{a_k}$ is determined by $\lceil \frac{R^{a_k}}{o_v^{a_k}} \rceil$, after which the Gurobi optimizer\cite{gurobi2023} generates a potentially infeasible deployment scheme. Finally, the migration method yields a feasible solution satisfying all imposed constraints.
	
\end{itemize}
\subsection{Experimental Results}
\begin{figure}[ht!]
	\centering
	\includegraphics[width=0.35\textwidth]{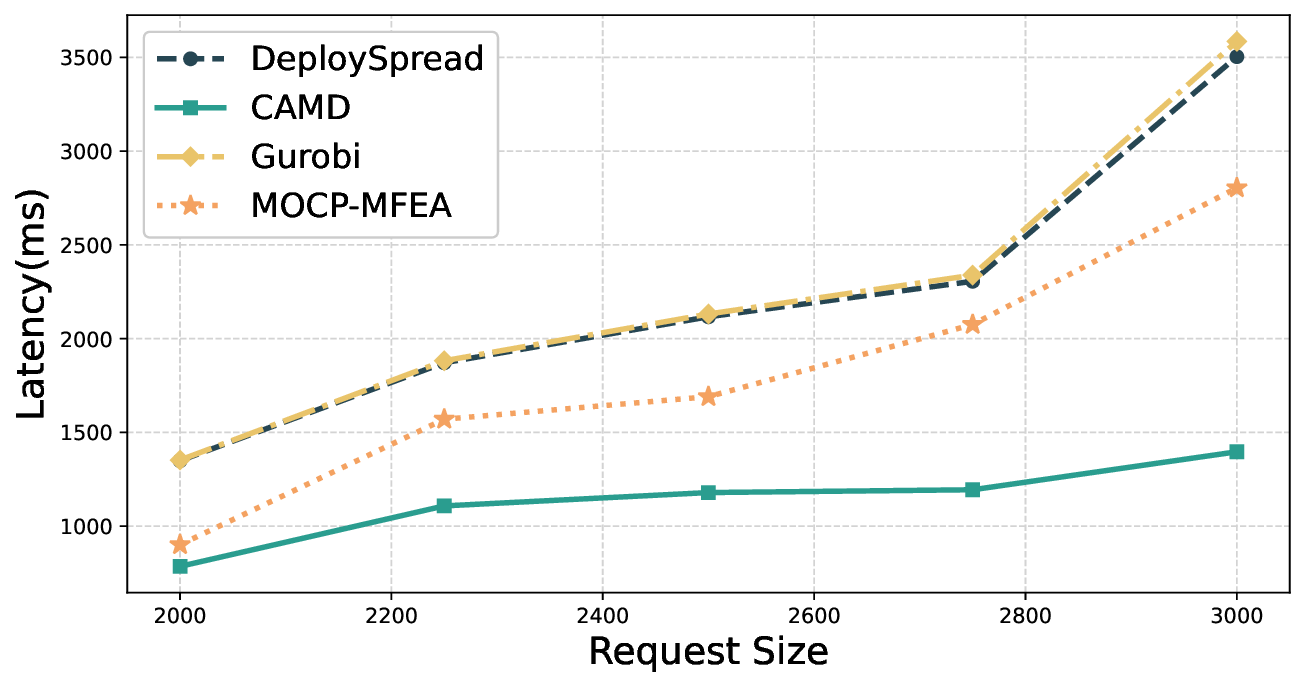}
	\caption{Response latency under various request sizes.}
	\label{fig:fig2}
\end{figure}

In Fig.~\ref{fig:fig2}, we show the performance of algorithms on response time with different numbers of user requests, ranging from 2000 to 3000. The system includes 3 servers and 3 types of applications, with each application consisting of 2-4 microservices. The response time increases as the number of user requests grows due to the limited computational resources available in the MEC system. Our proposed algorithm adapts well to the growing volume of user requests. Compared to baseline algorithms which show significantly increasing response latency, CAMD maintains a more stable performance as user load increases, highlighting its resource distribution capabilities in high-demand scenarios.

\begin{figure}[ht!]
	\centering
	\includegraphics[width=0.35\textwidth]{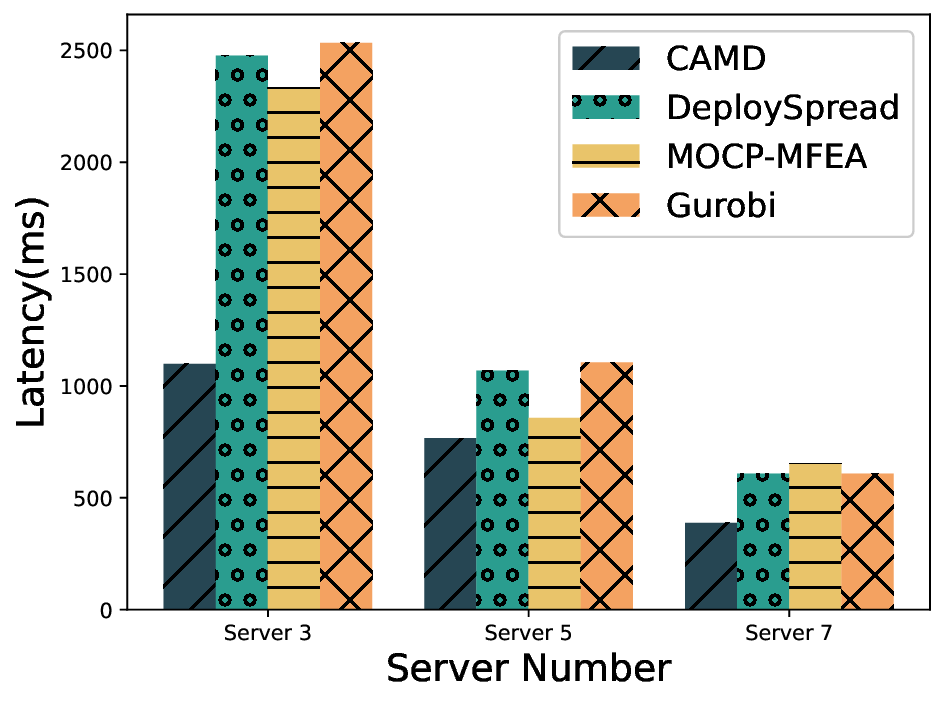}
	\caption{Response latency under various server numbers.}
	\label{fig:fig3}
\end{figure}

To evaluate the performance of algorithms at different server scales,  experiments are conducted with 3, 5, and 7 servers. We set user requests to 2000 and application types to 3, with each application consisting of 5-7 microservices. As shown in Fig.~\ref{fig:fig3}, the proposed algorithm outperforms others in reducing response latency, particularly in resource-limited environments. The key factor contributing to this outcome is that CAMD considers both request quantity and application priority, while baseline methods frequently overlook the latter aspect.

\begin{figure}[ht!]
	\centering
	\includegraphics[width=0.35\textwidth]{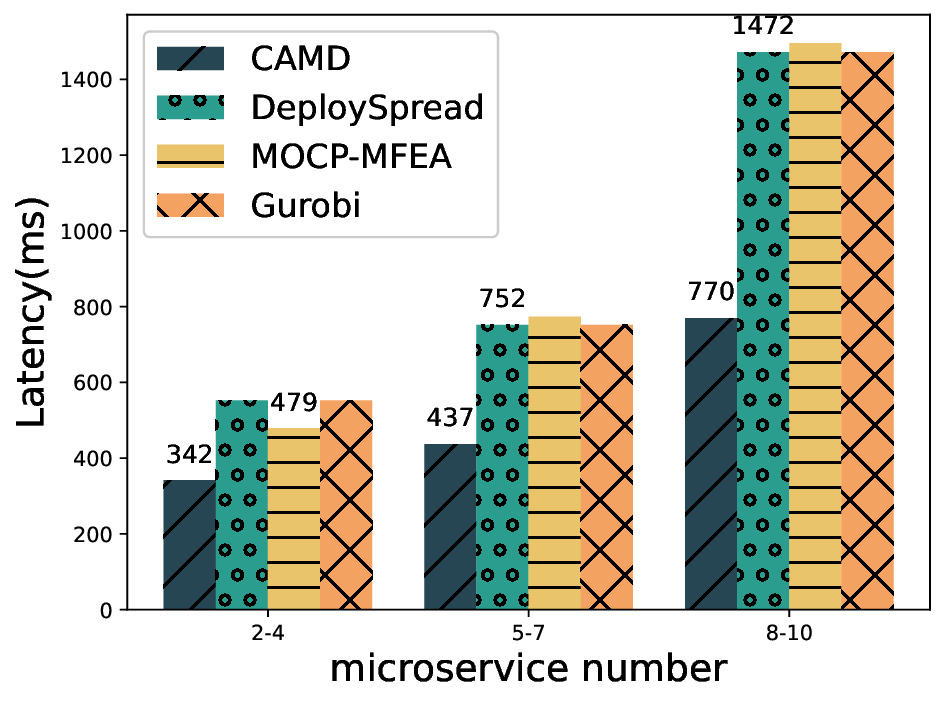}
	\caption{Response latency under various numbers of microservices composed of an application.}
	\label{fig:fig4}
\end{figure}

In Fig.~\ref{fig:fig4}, we change the number of microservices composed of an application to evaluate algorithm performance. In three sets of experiments, compared to the algorithm that performs best among baselines, the proposed algorithm results in latency reductions of 28.60\%, 41.89\%, and 47.69\% respectively. We can conclude that, for simpler applications, the proposed algorithm demonstrates clear advantages, reducing response delay compared to baselines. Furthermore, as application complexity increases, the superiority of the proposed algorithm becomes increasingly evident.
\vspace{-10pt} 
\begin{figure}
	\centering
	\includegraphics[width=0.5\textwidth]{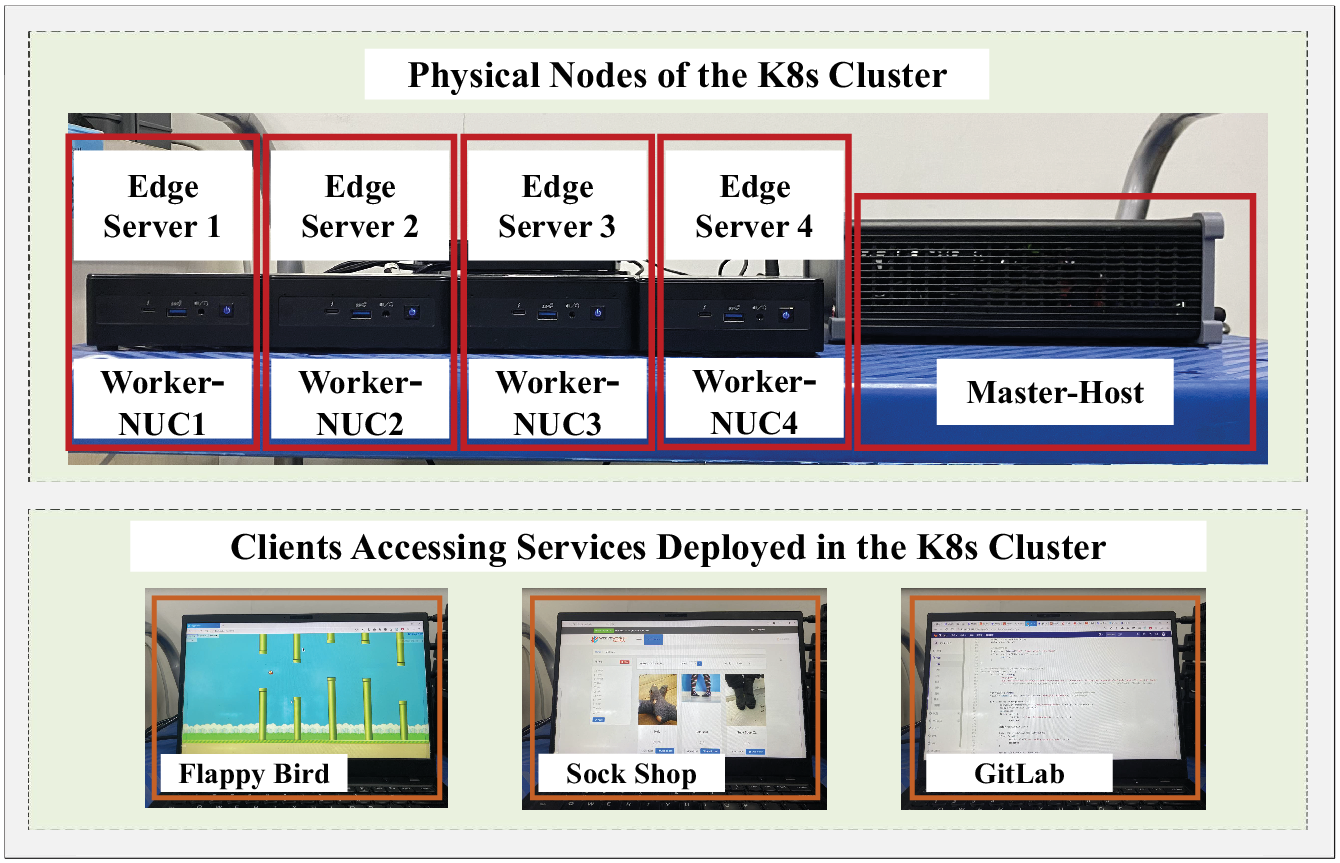}
	\caption{Collaborative MEC testbed for microservice deployment.}
	\label{fig:testbed}
\end{figure}
\subsection{Practicability Evaluation}
In this section, we construct a Kubernetes-based testbed to evaluate the practicality of the proposed algorithm. In our testbed, four Intel NUC mini computers act as edge servers, and the Host serves as a control platform. The testbed implements a Kubernetes (K8s) cluster, consisting of one master and four worker nodes. The master node corresponds to the Host, while each worker node corresponds to a NUC, as illustrated in Fig. ~\ref{fig:testbed}. While the master node can handle both management and application workloads, we opt to deploy it separately on the Host, solely for cluster management, to enhance stability, security, and performance. Prometheus is used to collect the current CPU and memory usage of each node. Within the K8s cluster, we deploy three services with different latency requirements, namely a Flappy Bird cloud game, Sock Shop, and GitLab. Of these services, cloud game demands the lowest latency, as any lag significantly impacts user experience. As an e-commerce application, Sock Shop requires moderate responsiveness, while GitLab can tolerate longer loading times.

\begin{figure}[ht!]
	\centering
	\includegraphics[width=0.35\textwidth]{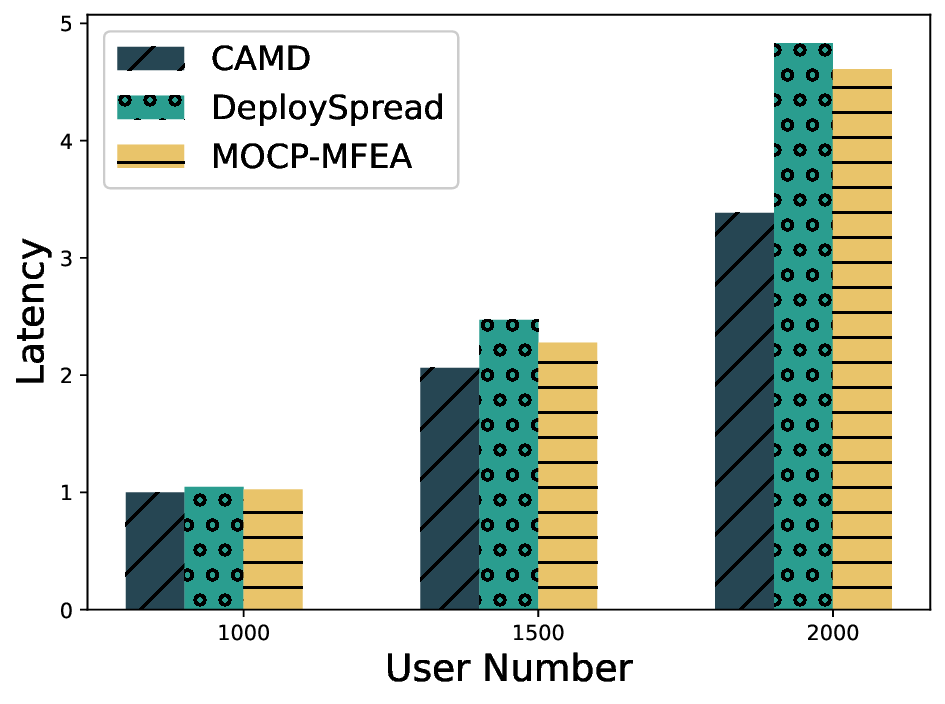}
	\caption{Response latency under various user numbers in the testbed.}
	\label{fig:fig5}
\end{figure}

In this testbed, we evaluate the performance of algorithms under concurrent user loads of 1000, 1500, and 2000 users. We normalize the latency results by setting the latency of CAMD under the 1000-concurrent-user scenario as the baseline (value of 1) and express the results for other scenarios as ratios relative to this baseline. As can be seen from Fig.~\ref{fig:fig5}, the proposed algorithm can reasonably allocate limited computing resources in the test platform and still has better performance in latency optimization. The results from the test platform are consistent with the simulation results presented in Fig. ~\ref{fig:fig2}, further validating the effectiveness and reliability of the proposed algorithm in practical scenarios. 

\section{Conclusion}
This paper studies the problem of microservice deployment of multiple applications in collaborative MEC under resource competition. We design the CAMD algorithm based on the BCD concept that decomposes the deployment problem into subproblems, subsequently employing a heuristic algorithm for each subproblem. Finally, the experimental findings indicate that our proposed algorithm allocates resources more efficiently and outperforms existing solutions in reducing response time.

\bibliographystyle{IEEEtran}
\bibliography{reference}

\begin{thebibliography}{10}
\providecommand{\url}[1]{#1}
\csname url@samestyle\endcsname
\providecommand{\newblock}{\relax}
\providecommand{\bibinfo}[2]{#2}
\providecommand{\BIBentrySTDinterwordspacing}{\spaceskip=0pt\relax}
\providecommand{\BIBentryALTinterwordstretchfactor}{4}
\providecommand{\BIBentryALTinterwordspacing}{\spaceskip=\fontdimen2\font plus
\BIBentryALTinterwordstretchfactor\fontdimen3\font minus
  \fontdimen4\font\relax}
\providecommand{\BIBforeignlanguage}[2]{{%
\expandafter\ifx\csname l@#1\endcsname\relax
\typeout{** WARNING: IEEEtran.bst: No hyphenation pattern has been}%
\typeout{** loaded for the language `#1'. Using the pattern for}%
\typeout{** the default language instead.}%
\else
\language=\csname l@#1\endcsname
\fi
#2}}
\providecommand{\BIBdecl}{\relax}
\BIBdecl

\bibitem{yukun2024computing}
S.~Y. et~al., ``Computing power network: A survey,'' \emph{China Commun.},
  vol.~21, no.~9, pp. 109--145, Sept.2024.

\bibitem{jamshidi2018microservices}
P.~Jamshidi, C.~Pahl, N.~C. Mendon{\c{c}}a, J.~Lewis, and S.~Tilkov,
  ``Microservices: The journey so far and challenges ahead,'' \emph{IEEE
  Softw.}, vol.~35, no.~3, pp. 24--35, 2018.

\bibitem{alqaisi2023containerized}
O.~I. Alqaisi, A.~{\c{S}}. Tosun, and T.~Korkmaz, ``Containerized computer
  vision applications on edge devices,'' in \emph{Proc. IEEE Int. Conf. Edge
  Comput. Commun. (EDGE)}, Chicago,USA, 2023, pp. 1--11.

\bibitem{li2024priority}
Y.~Li, X.~Zhang, B.~Lei, Z.~Qu, and W.~Wang, ``Priority and stackelberg
  game-based incentive task allocation for device-assisted mec networks,'' in
  \emph{Proc. IEEE Glob. Commun. Conf. (GLOBECOM)}, Cape Town, South Africa,
  2024.

\bibitem{qi2023edge}
J.~Qi, H.~Zhang, X.~Li, H.~Ji, and X.~Shao, ``Edge-edge collaboration based
  micro-service deployment in edge computing networks,'' in \emph{Proc. IEEE
  Wireless Commun. Netw. Conf. (WCNC)}, Glasgow, United Kingdom, 2023, pp.
  1--6.

\bibitem{zhou2023dependency}
J.~Zhou, G.~Wang, and W.~Zhou, ``Dependency-aware microservice deployment and
  resource allocation in distributed edge networks,'' in \emph{Proc. Int.
  Wireless Commun. Mobile Comput. Conf. (IWCMC)}, Marrakesh, Morocco, 2023, pp.
  568--573.

\bibitem{li2023topology}
X.~Li, J.~Zhou, X.~Wei, D.~Li, Z.~Qian, J.~Wu, X.~Qin, and S.~Lu,
  ``Topology-aware scheduling framework for microservice applications in
  cloud,'' \emph{IEEE Trans. Parallel Distrib. Syst.}, vol.~34, no.~5, pp.
  1635--1649, 2023.

\bibitem{deng2020optimal}
S.~D. et~al., ``Optimal application deployment in resource constrained
  distributed edges,'' \emph{IEEE Trans. Mobile Comput.}, vol.~20, no.~5, pp.
  1907--1923, 2020.

\bibitem{li2022application}
H.~Li, B.~Tang, W.~Xu, F.~Guo, and X.~Zhang, ``Application deployment in mobile
  edge computing environment based on microservice chain,'' in \emph{Proc. IEEE
  25th Int. Conf. Comput. Supported Cooperative Work Des.}, Hangzhou, China,
  2022, pp. 531--536.

\bibitem{wang2022online}
J.~Wang, J.~Hu, G.~Min, Q.~Ni, and T.~El-Ghazawi, ``Online service migration in
  mobile edge with incomplete system information: A deep recurrent actor-critic
  learning approach,'' \emph{IEEE Trans. Mobile Comput.}, vol.~22, no.~11, pp.
  6663--6675, 2022.

\bibitem{chen2023msm}
W.~Chen, M.~Liu, F.~Wu, H.~Wu, Y.~Miao, F.~Lyu, and X.~Shen, ``Msm:
  Mobility-aware service migration for seamless provision: A data-driven
  approach,'' \emph{IEEE Internet Things J.}, vol.~10, no.~17, pp.
  15\,690--15\,704, 2023.

\bibitem{li2023joint}
Y.~Li, X.~Ge, B.~Lei, X.~Zhang, and W.~Wang, ``Joint task partitioning and
  parallel scheduling in device-assisted mobile edge networks,'' \emph{IEEE
  Internet of Things Journal}, 2023.

\bibitem{he2022online}
X.~He, Z.~Tu, M.~Wagner, X.~Xu, and Z.~Wang, ``Online deployment algorithms for
  microservice systems with complex dependencies,'' \emph{IEEE Trans. Cloud
  Comput.}, vol.~11, no.~2, pp. 1746--1763, 2022.

\bibitem{liu2021multi}
R.~Liu, P.~Yang, H.~Lv, and W.~Li, ``Multi-objective multi-factorial
  evolutionary algorithm for container placement,'' \emph{IEEE Trans. Cloud
  Comput.}, vol.~11, no.~2, pp. 1430--1445, 2021.

\bibitem{gurobi2023}
\BIBentryALTinterwordspacing
\emph{"Gurobi Optimizer Reference Manual"}, 2023, [Online;. [Online].
  Available: \url{https://www.gurobi.com}
\BIBentrySTDinterwordspacing

\end{thebibliography}

\end{document}